\documentclass{article}
\usepackage{graphicx} 
\usepackage{amsmath,amssymb,amsfonts,amsthm}
\usepackage{makeidx}
\usepackage{xcolor}
\usepackage{subcaption}
\usepackage{cite}
\usepackage{hyperref}
\usepackage{tabularx}
\usepackage{booktabs}
\usepackage{array}
\usepackage{multirow}
\usepackage{longtable}
\usepackage{listings}

\lstdefinelanguage{json}{
    basicstyle=\ttfamily\small,
    numbers=left,
    numberstyle=\tiny\color{gray},
    stepnumber=1,
    numbersep=8pt,
    showstringspaces=false,
    breaklines=true,
    frame=single,
    backgroundcolor=\color{gray!5},
    string=[s]{"}{"},
    comment=[l]{//}
}

\newcommand{\eqn}{equation}
\newcommand{\lb}{\left(}
\newcommand{\rb}{\right)}
\newcommand{\GeV}{\text{GeV}}
\newcommand{\packagename}{TRSMScans}

\title{TRSMScans}
\author{Tania Robens$^1$,  Roxana Rodriguez$^2$, Manuel Samaniego$^2$ and Jason Veatch$^2$}
\date{
$^1$ Rudjer Boskovic Institute, Bijenicka cesta 54, 10000 Zagreb, Croatia\\
$^2$ California State University Stanislaus, 1 University Circle, 95382 Turlock, California, USA\\
\texttt{jason.veatch@cern.ch} \\
\texttt{trobens@irb.hr}
}
\begin{document}
\maketitle
\begin{abstract}
    In this work, we propose a new scan tool that automatically calculates maximal cross section predictions for a new physics scenario with additional scalar states.
    While the tool is currently optimized for asymmetric production and decay processes in the form of $p\,p\,\rightarrow\, h_3\,\rightarrow\,h_1\,h_2$, where $h_i$ denote CP even neutral scalars, in principle it can be extended to include any production and decay mode.
    As the code builds largely on the structure of the publicly available code \texttt{ScannerS}, it can in principle be extended to any other model contained within this tool.
    We here show first applications for a specific model, the Two-Real-Singlet Model (TRSM), and compare to publicly available results from the LHC collaborations from the previous runs.
    We also in detail discuss the structure of the code, including code installation, usage, as well as underlying algorithms. \\
    RBI-ThPhys-2026-26
\end{abstract}
\newpage

\section{Introduction}
\label{sec:intro}

The experimental collaborations at the LHC are pursuing a large
search program for Beyond the Standard Model (BSM) models with
extended scalar sectors.
These include processes with pair-production of novel scalars,
as well as single production via VBF or gluon-boson fusion,
scalar-strahlung from Drell-Yan production, and subsequent decays
into SM or new physics final states.
Many of these studies are performed in a model-independent way, with
assumptions about the kinematics and masses of the new particles, but
no other assumptions, such as coupling strengths.
Searches typically produce results in the form of upper limits on
production cross-section times branching ratio into the chosen final
state.
While model-independent searches are powerful tools for testing the
validity of the Standard Model and providing the potential to discover
new particles, it is interesting to compare the results to predictions
of specific BSM models.
Doing so makes it possible to understand whether the experimental searches
are sensitive to any region of the model's parameter space and whether
the limits can be used to constrain the model.

Many BSM models have multiple free parameters in the potential, making
a complete description of the maximum rates challenging. Conventionally,
experimental search results are presented in two-dimensional planes with
all other free parameters fixed. This can result in critical areas of
parameter space that the search may be sensitive to being missed.

In this work, we present a new tool \texttt{\packagename} that allows
the maximum predicted rates (cross-section times branching ratio)
predicted by a model for certain collider signatures to be found
efficiently with minimal human intervention.
The tool takes into account all current theoretical and experimental
constraints and scans over all free parameters of the models.
It largely relies on the publicly available tools \texttt{ScannerS}~\cite{Coimbra:2013gya,Ferreira:2014dya,Costa:2016afo,Muehlleitner:2016uwq,Muehlleitner:2020vzs} and \texttt{HiggsTools}~\cite{Bahl:2022higgstools},
building on \texttt{HiggsBounds}~\cite{Bechtle:2009xe,Bechtle:2011xm,Bechtle:2013hja,Bechtle:2013wla,Bechtle:2015nua,Bechtle:2020pkv,Bahl:2022exs} and \texttt{HiggsSignals}~\cite{Bechtle:2013xla,Stal:2013eps,Bechtle:2014pma,Bechtle:2020uop} 
\footnote{Note that the latest \texttt{ScannerS} version currently
does not provide a viable interface to \texttt{HiggsTools}. We are
therefore using a standalone interface in our work.}.
While the tool is currently optimized for the Two-Real-Singlet Model
(TRSM) \cite{Robens:2019kga,Robens:2022nnw}, it in principle allows for an automated
interface for all models that are available within these codes. 

This paper is structured as follows. Section~\ref{sec:parameters} introduces
the challenge of finding maximum rates in multi-dimensional parameter space.
Section~\ref{sec:algorithm} describes the optimization algorithm and shows
examples of its performance. Section~\ref{sec:code} describes the code in
detail, including installation, running optimizations, and plotting results.
Section~\ref{sec:cases} describes the Two-Real-Singlet-Model (TRSM) \cite{Robens:2019kga,Robens:2022nnw} as a use-case model and presents
select experimental results in the context of it.
Section~\ref{sec:conclusion} summarizes the paper and describes expected
future development of the tool.

\section{Parameter spaces}
\label{sec:parameters}

Many extended scalar models include many free parameters that
describe the potential and the additional particles. For example,
as discussed in Section~\ref{sec:trsm}, the TRSM has seven free
parameters. Two of these are masses of new scalars while the other
five are mixing angles and vacuum expectiation values (vevs). Experimental searches generally
define signal ``mass points'' for which the new particle masses
are fixed and the search is optimized for the resulting kinematics.
Following this philosophy, the mixing angles and vevs remain as free parameters for each mass point. For
each mass point, it is necessary to scan the free parameters over
their full allowed ranges to find the maximum predicted rates.

The rates for a given model can depend heavily on the exact values
of each of the free parameters. While the rates can be calculated for
any given set of parameter values, analytically finding the maximum
rates is not feasible. A common approach to finding maximum rates is
to randomly or systematically sample points in the parameter space and
take the highest sampled rate as an approximation of the highest allowed
rate. The challenge of this approach is the fact that for $N$ free
parameters, the sampling volume $V$ is described as
\begin{equation*}
    V = \prod_i^N R_i
\end{equation*}
where $R_i$ is the allowed range of parameter $i$. Conversely, the
sampled point density is inversely proportional to $V$. As the
number of free parameters increases, an increasing number of sampled
points is required to probe features in the parameter space and
converge on points close to the maximum allowed rates with a
reasonable degree of confidence. This is the fundamental challenge
posed by the problem of sampling $N$ dimensional space to find a
maximum. Since it is not possible to sample parameter spaces to an
arbitrary density and therefore precision, it is necessary to use
more intelligent algorithms, such as the tool presented in this paper.

The TRSM, which is used to show the tool's performance and usage, is
described in more detail in Section~\ref{sec:trsm}. The following is
a brief introduction to provide context for the performance demonstrations.

The TRSM introduces two additional real scalars to the SM scalar potential.
Two discrete symmetries are applied to minimize the number of additional
free parameters.
These are softly broken by the vevs of the additional fields, such that
all three gauge eigenstates mix to render physical mass states.
One the of novel states is identified as the SM-like Higgs measured by
the LHC experiments.
Similarly, the vev of the doublet is fixed by electroweak precision measurements.
This leaves a model with in total 7 free parameters:

\begin{\eqn}\label{eq:pars}
M_1,\,M_2,\,M_3,\,v_S,\,v_X,\,\theta_{hS},\,\theta_{hX},\,\theta_{SX}
\end{\eqn}

where one of the three masses is fixed and where the $v$s and $\theta$s
denote vevs and mixing angles, respectively.
The tool presented here is used to find the maximal rate for a fixed
set of scalar masses, which reduces the number of free parameters for
each mass point to 5.

\section{Optimization Algorithm}
\label{sec:algorithm}

The tool \texttt{\packagename} makes use of the \texttt{ScannerS} and
\texttt{HiggsTools} to calculate allowed production rates for BSM models
predicting scalar boson production.
\texttt{ScannerS} calculates production cross-sections, decay branching
ratios, resonance widths and rescalings, as well as theoretical constraints such as boundedness from below and perturbativity and constraints from electroweak precision observables.  \texttt{HiggsTools} on the other hand is used
to apply direct search and signal strength constraints on the models parameter
space.
\texttt{\packagename} provides a wrapper for these tools that intelligently
uses the random sampling \texttt{ScannerS} to converge on a nearly maximal
value of the allowed production rates and the corresponding values of the
model's free parameters.
By iteratively shrinking the parameter spaces for sampling to ``zoom in''
on the maximal value(s) and making use of parallel processing capabilities,
the optimization is done efficiently with minimal intervention required.

A single optimization scan samples the free parameters for a model with
a user-defined set of scalar particle masses and decay mode.
A list of the supported decay modes is given in Appendix~\ref{app:decay}.
The target quantity is the rate, denoted as  $xb$, that is defined as the production
cross section times the branching ratio of the scalars into the specified
decay products.
Prior to the optimization scan, an initial ``prescan'' is performed, sampling
the full free parameter space to provide a rough estimate of the topology and
outline of the allowed region(s).
Each prescan depends only on the scalar masses and is independent of production
and decay modes.
If a prescan already exists for a given mass point, a new prescan will not
be produced unless specified by the user.
Following the prescan, an optimization algorithm is implemented to estimate
the maximum allowed rate.
The ``zoom'' optimization algorithm is presented below.
At the time of the writing, other optimization algorithms are under development
but are not fully supported.

Based on the output of the prescan, a kernel density estimation is used to
split the full parameter volume into smaller volumes, each containing a
separate ``bump'' in $xb$.
Each of these smaller volumes is used as the starting point for a separate
``optimizer'' that iteratively converges on the local maxmimum in its defined
volume.
At each iteration, an optimizer samples a decreasing volume with an
increasing point density until stopping conditions are met, as described
in Section~\ref{sec:algorithm-stopping}.
After every individual optimizer satisfies one or more of its stopping
conditions, the optimization procedure stops and the sampled point with
the highest $xb$ is taken as the maximal allowed rate for the mass point
and decay mode.

\subsection{Filters}
\label{sec:algorithm-filters}

In our scan, we first call \texttt{ScannerS} to obtain points that are allowed by all constraints implemented in this tool. After a point has passed \texttt{ScannerS}, we furthermore test its validity against current experimental constraints and findings using \texttt{HiggsTools}. Furthermore,
filters are applied to all sampled point to ensure they are allowed. Any
point that fails a filter requirement is excluded from the optimization
procedure. The first filter checks whether sampled points are excluded
based on previous experimental limits using \texttt{HiggsTools}.
The second filter puts an upper bound on the relative widths of the
scalars. If any scalar has a relative width above the limit, the
point fails the filter. The width limits can be configured, as
discussed in Section~\ref{sec:code-config}, and are set to 15\% by default.

\subsection{Stopping conditions}
\label{sec:algorithm-stopping}

Optimizing a function through sampling does not provide any mechanism to
guarantee a global maximum has been found in a finite amount of time.
Therefore, it is necessary to define stopping conditions to ensure that
a value that is reasonably close to the global maximum has been found
in a manageable amount of time.

In a single optimization procedure, each zoom optimizer iterates until at
least one of two sets of stopping conditions is met.
The first set consists of the local stopping conditions that test whether
an optimizer continues to sufficiently increase its local maximum across
iterations.
The second set consists of global stopping conditions that test whether
an optimizer's local maximum is comparable to the highest maximum of all
optimizers.
The strictness of the local stopping conditions is set by the user, with
higher precision levels resulting in greater sampling density, a higher
confidence the result is close to the theoretical maximum, and generally
higher values of $xb$.
An adaptive precision strategy can also be used, in which the strictness
of the stopping conditions, and therefore the precision of an optimization
is modified automatically based on the ratio of the current highest value
of $xb$ to a target value, typically an experimental limit.
If the maximum sampled value is smaller than 1\% of the target value or
if it is more than 20 times larger than the target value, the optimization
is terminated early to avoid wasteful CPU usage.

\subsection{Performance}
\label{sec:performance}

\begin{figure}[!htbp]
    \centering

    \begin{subfigure}{0.49\textwidth}
        \centering
        \includegraphics[width=\linewidth]{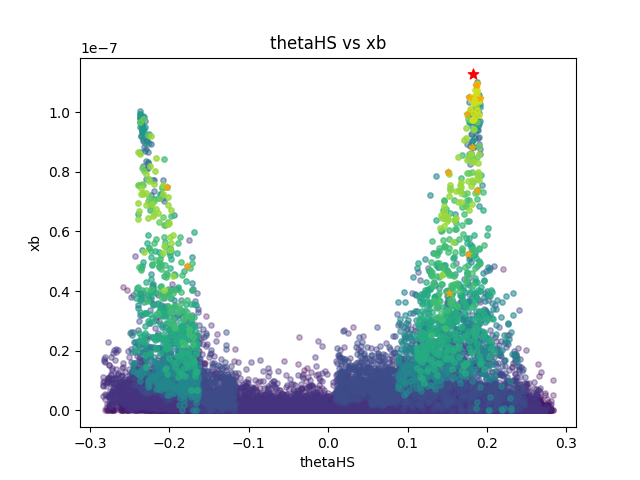}
        \caption{}
        \label{subfig:mass-point-xb-thetaHS}
    \end{subfigure}
    \hfill
    \begin{subfigure}{0.49\textwidth}
        \centering
        \includegraphics[width=\linewidth]{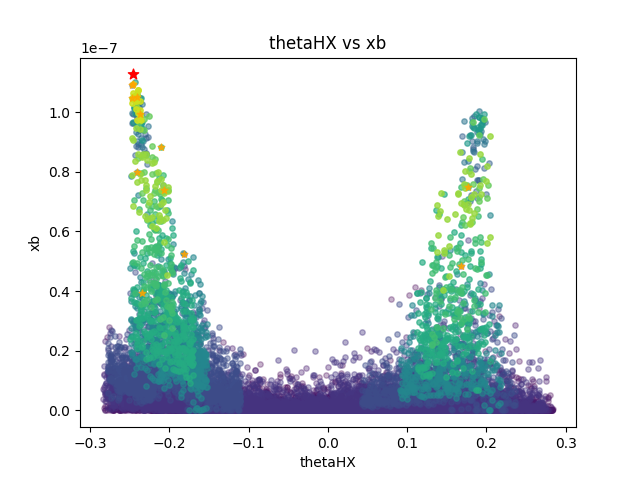}
        \caption{}
        \label{subfig:mass-point-xb-thetaHX}
    \end{subfigure}

    \begin{subfigure}{0.49\textwidth}
        \centering
        \includegraphics[width=\linewidth]{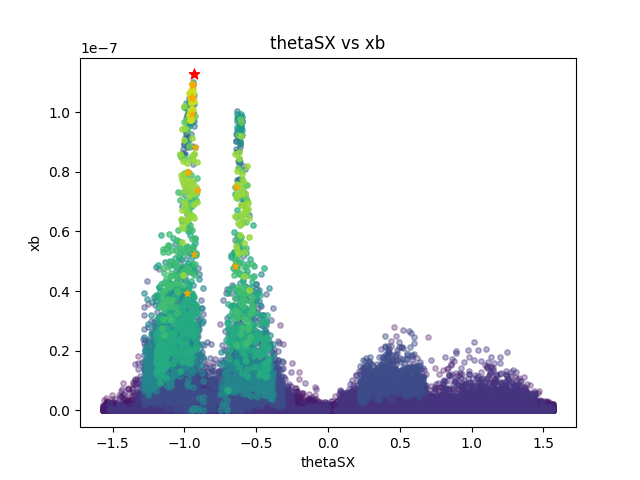}
        \caption{}
        \label{subfig:mass-point-xb-thetaSX}
    \end{subfigure}
    \caption{Evolution of the ``zoom'' optimization process for the $SH\rightarrow b\bar{b}\tau^+\tau^-$ final state with $m_X = 1000~\GeV$ and $m_S = 300~\GeV$ projected onto the three mixing angle parameters of the TRSM.
    The target quantity $xb$ is shown on the $y$-axis.
    The colors of the points show the scan iteration with purple being the prescan and subsequent iterations trending towards yellow.
    The small orange stars show each new maximum value of $xb$ found during the process and the red star indicates the overall maximum value of $xb$ found.}
    \label{fig:mass-point-xb-theta}
\end{figure}

\begin{figure}[!htbp]
    \centering
    \begin{subfigure}{0.49\textwidth}
        \centering
        \includegraphics[width=\linewidth]{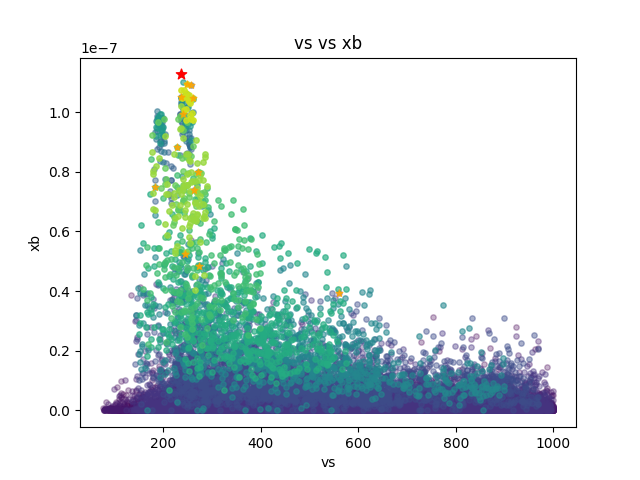}
        \caption{}
        \label{subfig:mass-point-xb-vs}
    \end{subfigure}
    \hfill
    \begin{subfigure}{0.49\textwidth}
        \centering
        \includegraphics[width=\linewidth]{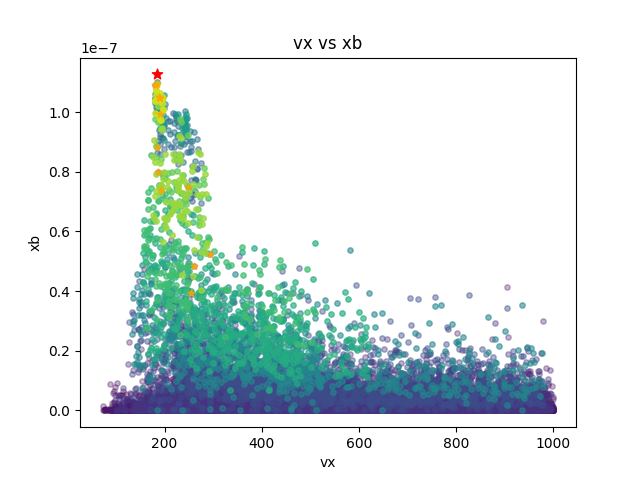}
        \caption{}
        \label{subfig:mass-point-xb-vx}
    \end{subfigure}
    \caption{Evolution of the ``zoom'' optimization process for the $SH\rightarrow b\bar{b}\tau^+\tau^-$ final state with $m_X = 1000~\GeV$ and $m_S = 300~\GeV$ projected onto the two vacuum expectation value parameters of the TRSM.
    The target quantity $xb$ is shown on the $y$-axis.
    The colors of the points show the scan iteration with purple being the prescan and subsequent iterations trending towards yellow.
    The small orange stars show each new maximum value of $xb$ found during the process and the red star indicates the overall maximum value of $xb$ found.}
    \label{fig:mass-point-xb-vev}
\end{figure}

\begin{figure}[!htbp]
    \centering

    \begin{subfigure}{0.49\textwidth}
        \centering
        \includegraphics[width=\linewidth]{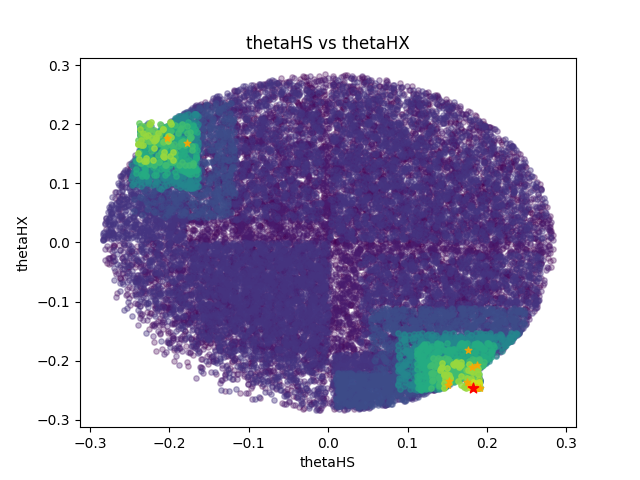}
        \caption{}
        \label{subfig:mass-point-thetaHS-thetaHX}
    \end{subfigure}
    \hfill
    \begin{subfigure}{0.49\textwidth}
        \centering
        \includegraphics[width=\linewidth]{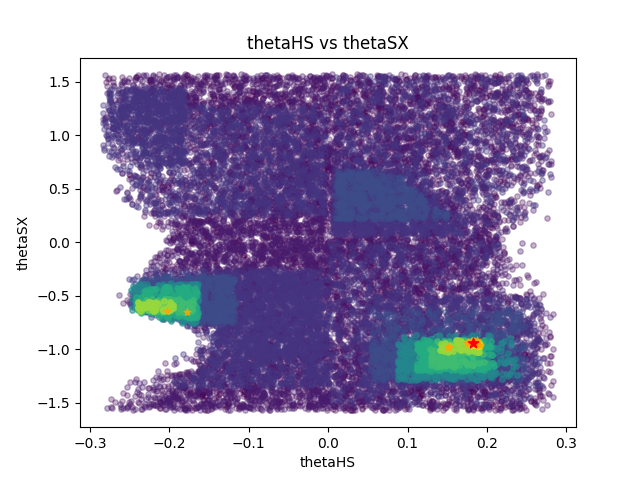}
        \caption{}
        \label{subfig:mass-point-thetaHS-thetaSX}
    \end{subfigure}

    \begin{subfigure}{0.49\textwidth}
        \centering
        \includegraphics[width=\linewidth]{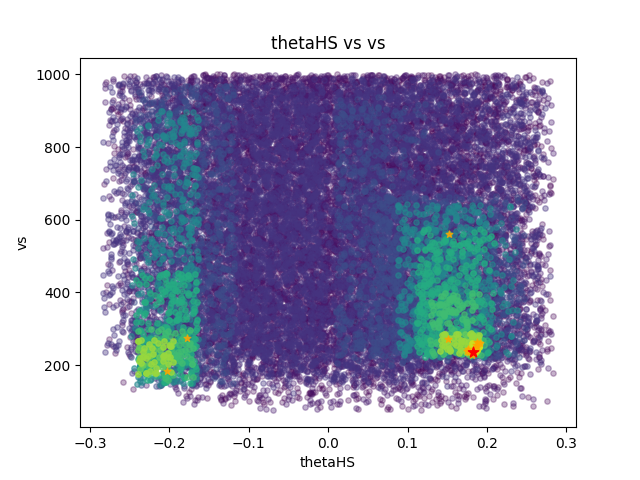}
        \caption{}
        \label{subfig:mass-point-thetaHS-vs}
    \end{subfigure}
    \hfill
    \begin{subfigure}{0.49\textwidth}
        \centering
        \includegraphics[width=\linewidth]{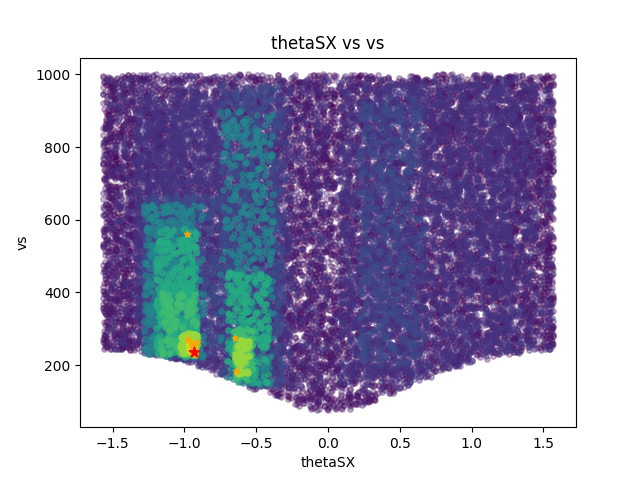}
        \caption{}
        \label{subfig:mass-point-thetaSX-vs}
    \end{subfigure}
    \caption{Evolution of the ``zoom'' optimization process for the $SH\rightarrow b\bar{b}\tau^+\tau^-$ final state with $m_X = 1000~\GeV$ and $m_S = 300~\GeV$ projected onto various pairs free parameters of the TRSM.
    The colors of the points show the scan iteration with purple being the prescan and subsequent iterations trending towards yellow.
    The small orange stars show each new maximum value of $xb$ found during the process and the red star indicates the overall maximum value of $xb$ found.}
    \label{fig:mass-point-2D}
\end{figure}

Figures~\ref{fig:mass-point-xb-theta}-\ref{fig:mass-point-2D} show the
evolution of sampled points through optimization iterations for an example
mass point for the TRSM model, described in Section~\ref{sec:trsm}.
Each plot shows the full set of sampled points, color-coded by iteration
number, with prescan points shown in dark blue and subsequent iterations
overlaid and transitioning to green and then to yellow.
The orange stars indicate the history of the maximum sampled $xb$ points
from one iteration to the next and the red star indicates the final maximum
point found.
Figure~\ref{fig:mass-point-xb-theta} shows $xb$ for all sampled points
projected onto single axes representing each of the three mixing angles.
It can be seen in each plot how multiple optimizers are concentrated around
local maxima and iteratively converge.
Figure~\ref{fig:mass-point-xb-vev} shows the same projected onto axes for
each of the vevs.
Figure~\ref{fig:mass-point-2D} shows projections of the sampled points
onto a selection of two-dimensional planes to demonstrate the regions
of parameter space in which each optimizer is concentrated.

\section{Code installation and usage}
\label{sec:code}

The \texttt{\packagename} tool is built on Python and is designed to have a
user-friendly interface and to include many useful tools.
It is publicly available at \href{https://github.com/jrveatch/trsmscans}{https://github.com/jrveatch/trsmscans}.
It has been designed for and tested on macOS and most common Linux distributions.
The tool can be installed and run on the CERN \texttt{lxplus} system, but
it is recommended to install your own version of Python rather than use
the default version.
Instructions for this can be found at~\cite{lxplus-python}.

\subsection{Pre-requisites and installation}
\label{sec:code-installation}

This sections outlines the minimum software requirements for the tools at
the time of this writing.
The requirements may change over time and will be reflected in the
repository README.md.

The minimum required Python version is 3.8 and can work with any version
available at the time of this writing.
In order to compile \texttt{ScannerS} and \texttt{HiggsTools}, it is
necessary to have compilers for \texttt{C}, \texttt{C++}, and Fortran.
The \texttt{C++} compiler must support \texttt{c++17}, which \texttt{gcc-9}
or newer does.
The version of \texttt{gcc} that ships with macOS does not include
\texttt{gfortran}, so a separate version of \texttt{gcc} needs to be
installed separately.
Other prerequisites include \texttt{gsl} (\texttt{libgsl-dev} on Ubuntu),
\texttt{Eigen3 3.3.0} or newer, and \texttt{clang 5} or newer.
Note that only version of \texttt{Eigen} that are 3.x are supported.
The newest versions of \texttt{Eigen} are 5.x and are not supported.
Most dependencies are checked automatically by the installation scripts
and all Python dependencies are installed automatically into a local
virtual environment.

In order to clone the repository and sub-repositories, accounts are needed
on both github\footnote{\href{https://github.com}{https://github.com}} and
gitlab\footnote{\href{https://gitlab.com}{https://gitlab.com}}.
Both SSH and https authentication are supported.
After \texttt{\packagename} is cloned, the command \verb|source setup.sh|
should be run from the top directory.
This will check many prerequisites, clone and compile the sub-repositories,
set necessary environment variables, and set up the Python virtual environment.
After the initial installation, \verb|source setup.sh| should be called at
the beginning of every new session to set environment variables and  activate
the virtual environment.
Calling the setup script with the \texttt{-f} option will force recompilation
and the \texttt{-c} option will trigger a clean setup that recreates all
setting and compiles from scratch.

\subsection{Running an optimization}

All optimization jobs are performed in the \texttt{run} directory using the
\texttt{launchscan.py} script, which is the primary entry point for running
the optimization procedure.
The script has many options that can be configured through the Command-Line Interface
(CLI), some of which are elaborated in this text, all of which are presented in
Appendix~\ref{app:cli}.
An example of how to run a basic optimization scan using the zoom optimization
for a single mass point is in \texttt{run/scan\_example\_zoom.sh}.
This script uses mass values that are unlikely to be used in an analysis
and a small number of sampling points to enable a quick functionality
verification.

The six necessary arguments for an optimization run for a single mass point
are shown in Table~\ref{tab:cli-basic-required}. These are used to specify
the BSM model, the BSM scalar masses, the decay mode, and the precision
of the scan.
Four optional, but useful arguments for a single mass point optimization are
shown in Table~\ref{tab:cli-basic-optional}.
The complete list of supported decay modes is given in Appendix~\ref{app:decay}.

\begin{table}[htbp!]
    \centering
    \caption{Minimum set of command-line arguments for a single mass point optimization.}
    \label{tab:cli-basic-required}
    \begin{tabular}{ll}
        \toprule
        \textbf{Argument} & \textbf{Description} \\
        \midrule
        \texttt{--mode} & Selects the operation mode: \texttt{scan} or \texttt{prescan}. \texttt{scan} triggers a prescan. \\
        \texttt{-X, --XMass} & Heavy scalar mass $m_X$ in GeV. \\
        \texttt{-S, --SMass} & Scalar mass $m_S$ in GeV. \\
        \texttt{-m, --model} & Physics model to scan. Defaults to \texttt{TRSMBroken}. \\
        \texttt{-d, --decay} & Decay mode used for the scan. \\
        \texttt{-p, --precision} & Level of precision for stopping conditions: \texttt{coarse}, \texttt{low}, \texttt{medium}, or \texttt{high} \\
        \bottomrule
    \end{tabular}
\end{table}

\begin{table}[htbp!]
    \centering
    \caption{Useful optional command-line arguments for a single mass point optimization.}
    \label{tab:cli-basic-optional}
    \begin{tabular}{ll}
        \toprule
        \textbf{Argument} & \textbf{Description} \\
        \midrule
        \texttt{-H, --HMass} & Scalar mass $m_H$ in GeV. Defaults to 125.09. \\
        \texttt{-s, --strategy} & Optimization strategy. Defaults to \texttt{zoom}. \\
        \texttt{-n, --num-points} & Initial number of scan points. Defaults to 10,000 for TRSM. \\
        \texttt{-f, --force-rerun} & Overwrite previous results. \\
        \bottomrule
    \end{tabular}
\end{table}

\subsection{Using a mass list}

In most scenarios, it is necessary to optimize a list of mass points
rather than a single mass point. While it is possible to run optimizations
for many mass points individually, this becomes logistically challenging
as the number increases. Instead, it is possible to define a full list
of mass points to be run over, either sequentially or in parallel (see
Section~\ref{sec:code-parallelization}). Mass lists are written as
\texttt{JSON} files in the \texttt{data/mass\_points} directory. An example
of a mass list \texttt{JSON} file is given in Appendix~\ref{app:mass_list}.
Tables~\ref{tab:cli-list-required} shows the five required CLI arguments
for a mass list optimization and Table~\ref{tab:cli-list-optional} shows
four optional, but useful, CLI arguments for a mass list optimization.

\begin{table}[htbp!]
    \centering
    \caption{Minimum set of command-line arguments for optimizing from a mass point list.}
    \label{tab:cli-list-required}
    \begin{tabular}{ll}
        \toprule
        \textbf{Argument} & \textbf{Description} \\
        \midrule
        \texttt{--mode} & Selects the operation mode: \texttt{scan} or \texttt{prescan}. \texttt{scan} triggers a prescan. \\
        \texttt{-l} & Flag to indicate a mass list should be used. \\
        \texttt{-i, --identifier} & String used to identify mass list name e.g., \texttt{ATLAS} or \texttt{CMS\_boosted}. \\
        \texttt{-m, --model} & Physics model to scan. Defaults to \texttt{TRSMBroken}. \\
        \texttt{-d, --decay} & Decay mode used for the scan. Used in mass list \texttt{.json} file name. \\
        \bottomrule
    \end{tabular}
\end{table}

\begin{table}[htbp!]
    \centering
    \caption{Useful optional command-line arguments for optimizing from a mass point list.}
    \label{tab:cli-list-optional}
    \begin{tabular}{ll}
        \toprule
        \textbf{Argument} & \textbf{Description} \\
        \midrule
        \texttt{-s, --strategy} & Optimization strategy. Defaults to \texttt{zoom}. \\
        \texttt{-n, --num-points} & Initial number of scan points. Defaults to 10,000 for TRSM. \\
        \texttt{-p, --precision} & If not set, precision adapts based on target rates. \\
        \texttt{-f, --force-rerun} & Overwrite previous results. \\
        \bottomrule
    \end{tabular}
\end{table}

\subsection{Configurations}
\label{sec:code-config}

In addition to the CLI arguments that can easily be changed from one
optimization run to the next, parameters that can be set by the user
but should typically be left constant are defined using \texttt{YML}
files in the \texttt{config} directory. \texttt{RunConfig.yml} sets
several configurations related to the computing strategy.
\texttt{OptimizerConfig.yml} sets parameters for the optimization
algorithms, including defining values for different stopping condition
precision levels. Configurations specific to a model are in files
with the \texttt{<model>\_<config-id>.yml}, where \texttt{config-id}
defaults to \texttt{default} and can be specified through a CLI
argument. Model configuration files set the relative width
limits and default numbers of points to sample in the optimization
procedure.

\subsection{Parallelization and batch mode}
\label{sec:code-parallelization}

A single mass point optimization relies on CPU-intensive \texttt{ScannerS}
sampling. Since point sampling is a sequence of large number of
independent steps, the tool uses parallel processing to make use
of multiple cores.
This is done using the Python \texttt{multiprocessing} library.

When running an optimization over a list of mass points on a single
machine, each mass point is run sequentially. A batch system such as
CERN's \texttt{HTCondor} system can be used to run each mass point
in parallel. This is done by using the \texttt{-b} CLI flag. The
\texttt{job-length} argument can be used to specify the \texttt{HTCondor}
job length.

\subsection{Plotting and mass point combinations}
\label{sec:code-plotting}

Scripts are provided to produce plots for single mass point optimizations
as well as combining the optimization results for a full mass list.
The \texttt{make\_plots.py} script is used to make all plots. An example
of its use for a single mass point is in \texttt{run/plot\_example\_zoom.sh}.
Mass point plots such as those shown in
Figures~\ref{fig:mass-point-xb-theta}-\ref{fig:mass-point-2D} can be produced
for a single mass point or using a mass list.
Mass plane plots combining the results of many individual mass points, such
as those shown in Figures~\ref{fig:combination-bbtautau-CMS} and
\ref{fig:combination-bbgamgam-ATLAS} can be produced using a mass list.
If a mass list is provided, the plotting script produces mass point and
combination plots by default. The \texttt{--only} CLI argument can be used
to produce exclusively mass point plots or combination plots.
The complete list of available CLI arguments is given in Appendix~\ref{app:cli}.

When optimizing a mass list, it is necessary to combine the outputs of each
mass point into a single file for easy processing. This is done using the
\texttt{combine\_results.py} script. The full set of available CLI arguments
is given in Table~\ref{tab:cli-combine}.

\begin{table}[htbp!]
    \centering
    \caption{Available command-line arguments for the \texttt{combine\_results.py} script.}
    \label{tab:cli-combine}
    \begin{tabular}{ll}
        \toprule
        \textbf{Argument} & \textbf{Description} \\
        \midrule
        \texttt{-m, --model} & Physics model to scan. Defaults to \texttt{TRSMBroken}. \\
        \texttt{-d, --decay} & Decay mode used for the scan. \\
        \texttt{-i, --identifier} & String used to identify mass list name e.g., \texttt{ATLAS} or \texttt{CMS\_boosted}. \\
        \texttt{-s, --strategy} & Optimization strategy. Defaults to \texttt{zoom}. \\
        \bottomrule
    \end{tabular}
\end{table}

\subsection{Output}
\label{sec:code-output}

All outputs are produced in the \texttt{run/output/<model>} directory. The
prescan output for each mass point is saved in
\texttt{run/output/<model>/prescan/<mass-point>} and consists of
\texttt{<model>\_prescan.tsv} that contains the full output from
\texttt{ScannerS} with appended filter information. The scan output for
each mass point is saved in \texttt{run/output/<model>/scan/<decay>/<mass-point>}
and consists of \texttt{summary\_<strategy>\_<model>\_<decay>\_<mass-point>.tsv}
that contains the evolution of the best point found by the optimization procedure
from one iteration to the next. The last line in the final represents the
best point after the stopping conditions are met. Each row summarizes the crucial
information about the corresponding point, including the rate, the values of each
of the free parameters, rescalings, and scalar mass widths. The output from a
mass list combination is saved as
\texttt{run/output/<model>/combination/<decay>\_<identifier>\_combination.tsv}
and contains the optimal point from each mass point in a single file. The plot
outputs are saved in \texttt{run/output/<model>/plots/<decay>} under
\texttt{mass\_points} and \texttt{combination}.

\section{Case studies}
\label{sec:cases}

In principle, the \texttt{\packagename} tool can be used to find maximal
rates for any extended scalar model that is available in \texttt{ScannerS}.
However, the current version of the code focuses on the TRSM as the primary
use-case, but the model definitions can be expanded with relative ease to
incorporate the particle predictions and free parameters for other models.

\subsection{The TRSM - model description}
\label{sec:trsm}

The Two-Real-Singlet Extension of the SM is a model that extends the scalar
sector of the SM by two additional scalar fields $S,\,X$ that transform as
singlets under the $SU(2)\,\times\,U(1)$ gauge group of the SM.
In order to reduce the number of free parameters of the model, two discrete
symmetries are additionally imposed such that

\begin{equation*}
\mathbb{Z}_2: S\,\rightarrow\,-S,\;\mathbb{Z}'_2: X\,\rightarrow\,-X,
\end{equation*}
while all other fields transform evenly under the respective symmetries.
With the above requirement, the most general renormalizable potential of the model is given by
\begin{equation}
    \begin{aligned}
        V\lb \Phi,\,S,\,X\rb & = \mu_{\Phi}^2 \Phi^\dagger \Phi + \lambda_{\Phi} {(\Phi^\dagger\Phi)}^2
        + \mu_{S}^2 S^2 + \lambda_S S^4
        + \mu_{X}^2 X^2 + \lambda_X X^2                                              \\
          & \quad+ \lambda_{\Phi S} \Phi^\dagger \Phi S^2
        + \lambda_{\Phi X} \Phi^\dagger \Phi X^2
        + \lambda_{SX} S^2 X^2,
    \end{aligned}\label{eq:TRSMpot}
\end{equation}

where $\Phi$ now denotes the SM-like doublet that gives rise to electroweak
symmetry breaking.
In the scenario considered here, all three fields obtain a vacuum expectation
value such that

\begin{equation}
    \Phi = \begin{pmatrix} 0\\\frac{\phi_h + v}{\sqrt{2}}\end{pmatrix}\;,
    S = \frac{\phi_S + v_S}{\sqrt{2}}\;, \quad
    X = \frac{\phi_X + v_X}{\sqrt{2}}\;.
    \label{eq:fields}
\end{equation}

This parametrization then leads to a breaking of the above symmetries and
subsequently to mixing of the 3 fields.
The interactions of the mass eigenstates to SM particles are then inherited
from the SM-like doublet via mixing.

The general setup of the model, including nomenclature and theoretical and
experimental constraints, has been vastly discussed in \cite{Robens:2019kga,Robens:2022nnw} (see also \cite{Robens:2025tew} for
most recent updates) and will not be repeated here.
Instead we briefly summarize the most important features of the model in
the following

\begin{itemize}
\item{} After electroweak symmetry breaking, the model has in general 9 free parameters, which we choose as
\begin{\eqn}\label{eq:pars}
M_1,\,M_2,\,M_3,\,v,\,v_S,\,v_X,\,\theta_{hS},\,\theta_{hX},\,\theta_{SX}
\end{\eqn}
where $M_i$ denotes the mass of scalar $h_i$ in the mass-eigenbasis, $v,\,v_S,\,v_X$ are the vacuum expectation values of the three fields and $\theta_{ij}$ denote mixing angles that define the $3\,\times\,3$ rotation matrix.
We here furthermore imply a mass hierarchy
\begin{\eqn*}
M_1\,\leq\,M_2\,\leq\,M_3.
\end{\eqn*}
After electroweak symmetry breaking, two of the parameters in eqn (\ref{eq:pars}) are fixed, namely, the vev of the SM-like doublet $v\,\sim\,246\,\GeV$ as well as one of the three masses $M_i\,\sim\,125\,\GeV$.
Note that here $i$ can be either of the three masses leading to different phenomenological scenarios.
\item{}Due to the above mixing, there are three rescaling factors $\kappa_i$ that are used to rescale the couplings of the three scalar fields with respect to couplings of an SM-like scalar $h_i$ according to
\begin{\eqn}\label{eq:resc}
g_i\,=\,\kappa_i\,g^{\text SM}
\end{\eqn}
Note that these scaling factors are universal for all couplings to SM particles and also obey the important sum rule \cite{Gunion:1990kf}
\begin{\eqn}\label{eq:sumrule}
\sum_i\,\kappa_i^2\,=\,1.
\end{\eqn}
Therefore the widths of the above scalars is given by
\begin{\eqn*}
\Gamma^\text{tot}\lb h_i \rb\,=\,\kappa^2_i\,\Gamma^\text{SM}\lb h_i\rb\,+\,\sum_{j,k} \Gamma_{h_i\,\rightarrow\,h_j\,h_k}\lb h_i\rb
\end{\eqn*}
where the latter denotes the sum over the partial decay widths into the other scalars of the model.
Note that if the latter contribution vanishes, the branching ratios into SM final state are identical to the ones of the SM scalar of the same mass.
Plots denoting these branching ratios can e.g. be found in \cite{Robens:2019kga}.

Equation (\ref{eq:resc}) also leads to a rescaling of the respective production cross sections according to
\begin{\eqn*}
\sigma^\text{prod}\lb  h_i \rb\,=\,\kappa_i^2\,\sigma^\text{prod, SM}\lb h_i\rb
\end{\eqn*}
where again the latter quantity denotes the production cross section for an SM-like scalar $h_i$.
\item{} The TRSM contains 3 CP even neutral scalars that mix.
It therefore allows for all possible triple and quartic scalar interactions between the three scalars.
In particular, we can have the following decay chains
\begin{\eqn}\label{eq:decs}
h_i\,\rightarrow\,h_j h_j,\, h_i\rightarrow\,h_j h_k,
\end{\eqn}
where the actual branching ratios for the above processes depend on the specific points in parameter space.
Each scalar can then decay either into scalars as show above or into SM -like particles.
Due to the sum rule (\ref{eq:sumrule}), production of any of the scalars is suppressed with respect to the rates achievable for a SM-like scalar of the same mass for scalars with masses $\neq\,125\,\GeV$.
Current signal strength gives suppression factors $\mathcal{O}\lb 10\%\rb$.

\end{itemize}
Note that the decay modes in (\ref{eq:decs}) in particular also allow for
triple and quartic scalar final states.
For more details on the model, including current constraints and available
parameter regions, please refer to the literature above.

When using \texttt{\packagename}, the model name \texttt{TRSMBroken} should be
used for the TRSM.

\subsection{Application to experimental results}
\label{sec:analyses}

Applying the results of optimization scans to experimental results
consists of two steps. The first is to find the maximal rates for
each mass point investigated by an experimental search. The second
is to compare the maximal rates to the expected and observed experimental
limits. If the experimental limit is above the maximal rate at a mass
point, the search is not sensitive to the model at that mass point.
If the experimental limit is below the maximal rate at a mass point,
the search is sensitive to the model at that mass point and regions
of model parameter space can be excluded.

\begin{figure}[!htbp]
    \centering
    \begin{subfigure}{0.49\textwidth}
        \centering
        \includegraphics[width=\linewidth]{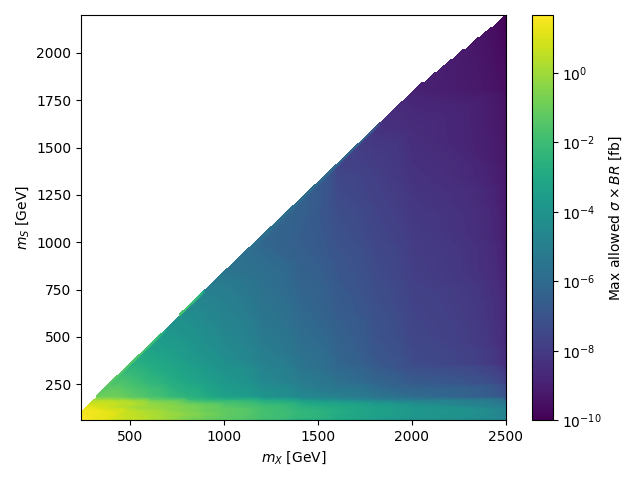}
        \caption{}
        \label{subfig:combination-bbtautau-CMS-xbmax}
    \end{subfigure}

    \begin{subfigure}{0.49\textwidth}
        \centering
        \includegraphics[width=\linewidth]{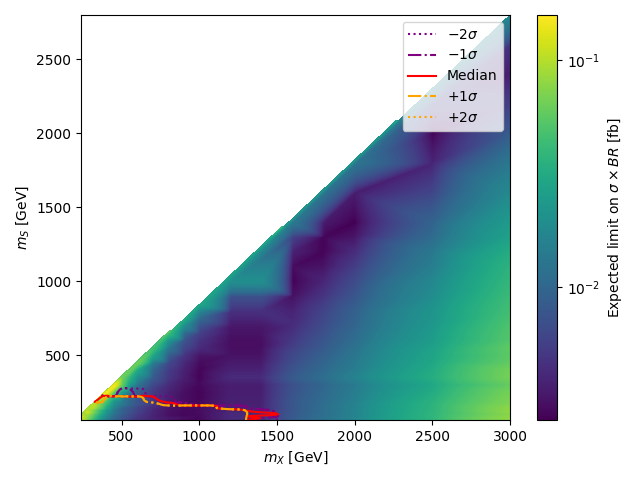}
        \caption{}
        \label{subfig:combination-bbtautau-CMS-expected}
    \end{subfigure}
    \hfill
    \begin{subfigure}{0.49\textwidth}
        \centering
        \includegraphics[width=\linewidth]{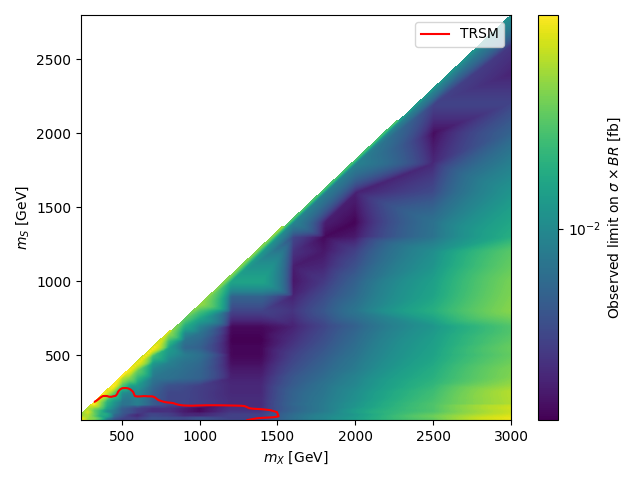}
        \caption{}
        \label{subfig:combination-bbtautau-CMS-observed}
    \end{subfigure}
    \caption{Summary plots of the maximum allowed $xb$ in the TRSM for a range of mass points in the $m_X$-$m_S$ plane.
    (\subref{subfig:combination-bbtautau-CMS-xbmax}) shows the maximum sampled values. 
    (\subref{subfig:combination-bbtautau-CMS-expected}) shows expected experimental limits and (\subref{subfig:combination-bbtautau-CMS-observed}) shows observed experimental limits.
    Both plots have contours indicating regions where the experimental limits are below the maximum allowed $xb$, such that constraints can be placed on the TRSM.
    (\subref{subfig:combination-bbtautau-CMS-expected}) includes contours for the nominal expected limits as well as the $\pm1\sigma$ and $\pm2\sigma$ variations.
    Experimental limit results are from the full Run 2 CMS $SH\rightarrow b\bar{b}\tau^+\tau^-$ analysis~\cite{CMS-HIG-20-014}.}
    \label{fig:combination-bbtautau-CMS}
\end{figure}

\begin{figure}[!htbp]
    \centering

    \begin{subfigure}{0.49\textwidth}
        \centering
        \includegraphics[width=\linewidth]{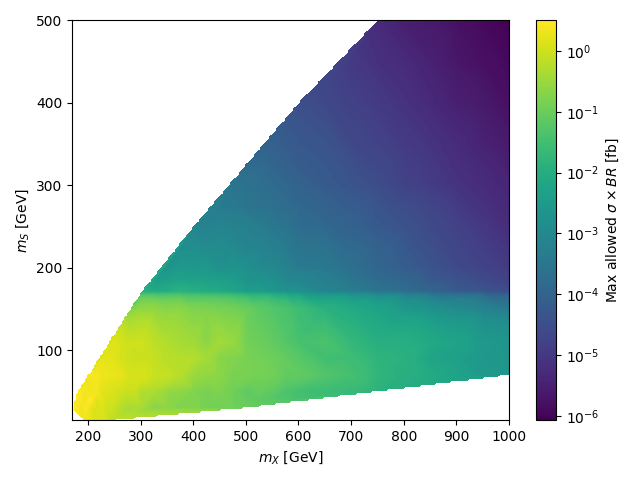}
        \caption{}
        \label{subfig:combination-bbgamgam-ATLAS-xbmax}
    \end{subfigure}

    \begin{subfigure}{0.49\textwidth}
        \centering
        \includegraphics[width=\linewidth]{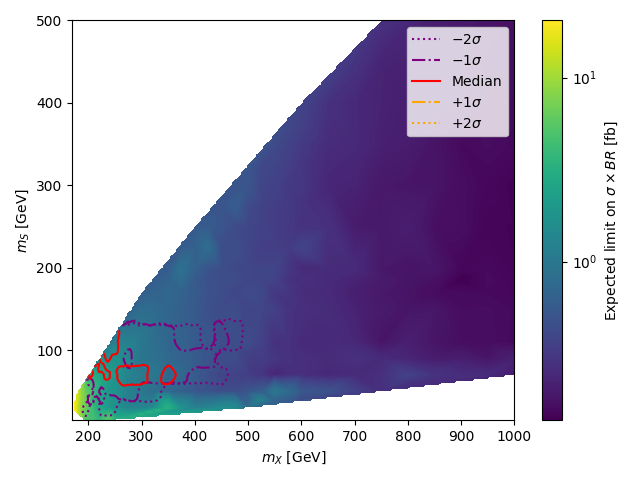}
        \caption{}
        \label{subfig:combination-bbgamgam-ATLAS-expected}
    \end{subfigure}
    \hfill
    \begin{subfigure}{0.49\textwidth}
        \centering
        \includegraphics[width=\linewidth]{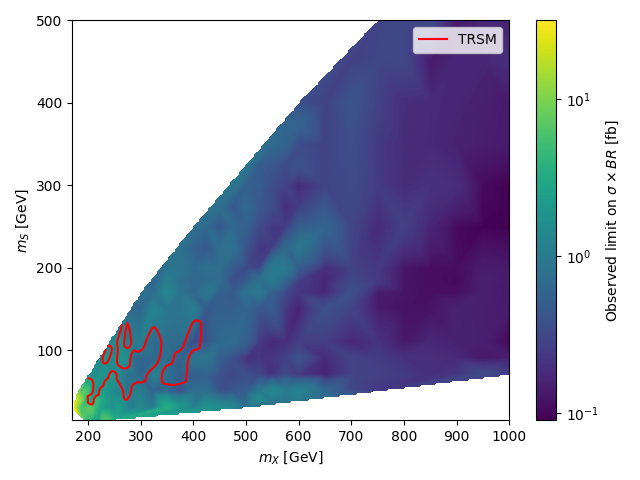}
        \caption{}
        \label{subfig:combination-bbgamgam-ATLAS-observed}
    \end{subfigure}
    \caption{Summary plots of the maximum allowed $xb$ in the TRSM for a range of mass points in the $m_X$-$m_S$ plane.
    (\subref{subfig:combination-bbgamgam-ATLAS-xbmax}) shows the maximum sampled values. 
    (\subref{subfig:combination-bbgamgam-ATLAS-expected}) shows expected experimental limits and (\subref{subfig:combination-bbgamgam-ATLAS-observed}) shows observed experimental limits.
    Both plots have contours indicating regions where the experimental limits are below the maximum allowed $xb$, such that constraints can be placed on the TRSM.
    (\subref{subfig:combination-bbgamgam-ATLAS-expected}) includes contours for the nominal expected limits as well as the $\pm1\sigma$ and $\pm2\sigma$ variations.
    Experimental limit results are from the full Run 2 ATLAS $SH\rightarrow b\bar{b}\gamma\gamma$ analysis~\cite{HDBS-2021-17}.}
    \label{fig:combination-bbgamgam-ATLAS}
\end{figure}

Figure~\ref{fig:combination-bbtautau-CMS} shows the $m_X$-$m_S$ plane
with maximal allowed TRSM rates as well as with experimental limits
and contours indicating regions sensitive to the TRSM for the CMS search
for $SH\rightarrow b\bar{b}\tau^+\tau^-$~\cite{CMS-HIG-20-014}.
The same plots are shown in Figure~\ref{fig:combination-bbgamgam-ATLAS}
for the ATLAS search for $SH\rightarrow b\bar{b}\gamma\gamma$~\cite{HDBS-2021-17}.

\section{Conclusions and outlook}
\label{sec:conclusion}

In this paper, we presented the \texttt{\packagename} tool that provides
a user-friendly method to find maximal cross-section times branching ratios
for extended Higgs sector models predicting additional scalar bosons.
The tool provides a Python interface that makes use of \texttt{ScannerS}
and \texttt{HiggsTools} to sample model parameter spaces efficiently.
The tool continues to be under development. Expected additions include
more sophisticated and faster optimization strategies, computational
optimization, and other quality of life improvements.

\newpage

\appendix

\section{CLI options}
\label{app:cli}

\begin{table}[htbp!]
\centering
\caption{Command-line options for \texttt{launch\_scan.py}.}
\label{tab:scan-options}
\begin{tabularx}{\linewidth}{llX}
\toprule
\textbf{Option} & \textbf{Default} & \textbf{Description} \\
\midrule
\texttt{--mode} & required & Selects the operation mode: \texttt{scan} or \texttt{prescan}. \\
\texttt{-b, --batch} & false & Submit jobs to HTCondor rather than running interactively. \\
\texttt{-l, --use-mass-list} & false & Run over a list of mass points instead of one specified point. \\
\texttt{-X, --XMass} & none & Heavy scalar mass $m_X$ in GeV. \\
\texttt{-S, --SMass} & none & Scalar mass $m_S$ in GeV. \\
\texttt{-H, --HMass} & 125.09 & Scalar mass $m_H$ in GeV. \\
\texttt{-i, --identifier} & none & Identifier used to label the selected mass set. \\
\texttt{-m, --model} & \texttt{TRSMBroken} & Physics model to scan. Must be one of the supported models. \\
\texttt{-c, --config-id} & \texttt{default} & Identifier for the model configuration file. \\
\texttt{-d, --decay} & none & Decay mode used when evaluating the scan. \\
\texttt{-s, --strategy} & \texttt{zoom} & Optimization strategy: \texttt{zoom}, \texttt{meanshift}, or \texttt{bayes}. \\
\texttt{-n, --num-points} & -1 & Initial number of scan points. A value of $-1$ uses the configured default. \\
\texttt{--limit-target} & none & Target limit used to determine the required precision dynamically. \\
\texttt{--prescan-points} & -1 & Number of prescan points to use when running in \texttt{scan} mode. \\
\texttt{-t, --iterations} & -1 & Maximum number of iterations or optimizers. A value of $-1$ uses the configured default. \\
\texttt{--num-cpus} & none & Number of CPUs requested for the job. \\
\texttt{-j, --job-length} & none & HTCondor job-length strategy. Must be one of the configured job lengths. \\
\texttt{--log-level} & \texttt{info} & Logging verbosity. \\
\texttt{--dry-run} & false & Print the submission commands without calling \texttt{condor\_submit}. \\
\texttt{-p, --precision} & automatic & Fix the optimization precision level instead of adapting it automatically. \\
\texttt{-r, --rerun-precision} & none & Rerun existing scan jobs if their precision is at or below this level. Ignored for prescans or when \texttt{--force-rerun} is used. \\
\texttt{-f, --force-rerun} & false & Force a rerun and overwrite previous results. \\
\bottomrule
\end{tabularx}
\end{table}

\begin{table}[htbp!]
\centering
\caption{Command-line options for \texttt{make\_plots.py}.}
\label{tab:plot-options}
\begin{tabularx}{\linewidth}{llX}
\toprule
\textbf{Option} & \textbf{Default} & \textbf{Description} \\
\midrule
\texttt{-l, --use-mass-list} & false & Run over a list of mass points instead of one specified point. \\
\texttt{-X, --XMass} & none & Heavy scalar mass $m_X$ in GeV. \\
\texttt{-S, --SMass} & none & Scalar mass $m_S$ in GeV. \\
\texttt{-H, --HMass} & 125.09 & Scalar mass $m_H$ in GeV. \\
\texttt{-i, --identifier} & none & Identifier used to label the selected mass set. \\
\texttt{-m, --model} & \texttt{TRSMBroken} & Physics model to scan. Must be one of the supported models. \\
\texttt{-d, --decay} & none & Decay mode used when evaluating the scan. \\
\texttt{-s, --strategy} & \texttt{zoom} & Optimization strategy: \texttt{zoom}, \texttt{meanshift}, or \texttt{bayes}. \\
\texttt{--no-sigma-bands} & none & Do not plot $\pm1\sigma$ and $\pm2\sigma$ expected limit contours. \\
\texttt{--no-plot-limits} & none & Do not produce exclusion limits plots. \\
\texttt{--log-mX} & none & Use logarithmic scale for the $m_X$ axis. \\
\texttt{--log-mS} & none & Use logarithmic scale for the $m_S$ axis. \\
\texttt{--log-axes} & none & Use logarithmic scale for both axes (equivalent to \texttt{--log-mX --log-mS}. \\
\texttt{--only} & none & Limit plotting to a single type: \texttt{masspoints} or \texttt{combination}. \\
\bottomrule
\end{tabularx}
\end{table}

\clearpage

\section{Supported production/decay modes}
\label{app:decay}

\begin{longtable}{
  >{\centering\arraybackslash}m{0.18\textwidth}
  >{\centering\arraybackslash}m{0.24\textwidth}
  >{\centering\arraybackslash}m{0.50\textwidth}
}
\caption{
Supported decay modes for resonant production in scalar models containing
$X$, $S$, and $H$, where $m_H = 125.09~\mathrm{GeV}$. Decay arguments
beginning with \texttt{SH} represent the sum of the rates of the two
possible assignments of the final-state decays to $S$ and $H$, which is
appropriate when the $S$ and $H$ contributions are treated inclusively.
Decay arguments beginning with \texttt{X} represent the inclusive sum
of all contributing $HH$, $SH$, and $SS$
production modes for the corresponding final state.
}
\label{tab:allowed-decay-modes}
\\
\toprule
Final state & Decay argument & Production/decay mode \\
\midrule
\endfirsthead

\toprule
Final state & Decay argument & Production/decay mode \\
\midrule
\endhead

\multirow{4}{*}{$b\bar{b}b\bar{b}$}
& $\mathtt{HHbbbb}$ & $X \rightarrow HH \rightarrow b\bar{b}b\bar{b}$ \\*
& $\mathtt{SHbbbb}$ & $X \rightarrow SH \rightarrow b\bar{b}b\bar{b}$ \\*
& $\mathtt{SSbbbb}$ & $X \rightarrow SS \rightarrow b\bar{b}b\bar{b}$ \\*
& $\mathtt{Xbbbb}$  & $X \rightarrow HH/SH/SS \rightarrow b\bar{b}b\bar{b}$ \\

\midrule
\multirow{6}{*}{$b\bar{b}\tau^+\tau^-$}
& $\mathtt{HHbbtautau}$ & $X \rightarrow HH \rightarrow b\bar{b}\tau^+\tau^-$ \\*
& $\mathtt{SbbHtautau}$ & $X \rightarrow S(b\bar{b})H(\tau^+\tau^-)$ \\*
& $\mathtt{StautauHbb}$ & $X \rightarrow S(\tau^+\tau^-)H(b\bar{b})$ \\*
& $\mathtt{SHbbtautau}$ & $X \rightarrow SH \rightarrow b\bar{b}\tau^+\tau^-$ \\*
& $\mathtt{SSbbtautau}$ & $X \rightarrow SS \rightarrow b\bar{b}\tau^+\tau^-$ \\*
& $\mathtt{Xbbtautau}$  & $X \rightarrow HH/SH/SS \rightarrow b\bar{b}\tau^+\tau^-$ \\

\midrule
\multirow{6}{*}{$b\bar{b}W^+W^-$}
& $\mathtt{HHbbWW}$ & $X \rightarrow HH \rightarrow b\bar{b}W^+W^-$ \\*
& $\mathtt{SbbHWW}$ & $X \rightarrow S(b\bar{b})H(W^+W^-)$ \\*
& $\mathtt{SWWHbb}$ & $X \rightarrow S(W^+W^-)H(b\bar{b})$ \\*
& $\mathtt{SHbbWW}$ & $X \rightarrow SH \rightarrow b\bar{b}W^+W^-$ \\*
& $\mathtt{SSbbWW}$ & $X \rightarrow SS \rightarrow b\bar{b}W^+W^-$ \\*
& $\mathtt{XbbWW}$  & $X \rightarrow HH/SH/SS \rightarrow b\bar{b}W^+W^-$ \\

\midrule
\multirow{6}{*}{$b\bar{b}ZZ$}
& $\mathtt{HHbbZZ}$ & $X \rightarrow HH \rightarrow b\bar{b}ZZ$ \\*
& $\mathtt{SbbHZZ}$ & $X \rightarrow S(b\bar{b})H(ZZ)$ \\*
& $\mathtt{SZZHbb}$ & $X \rightarrow S(ZZ)H(b\bar{b})$ \\*
& $\mathtt{SHbbZZ}$ & $X \rightarrow SH \rightarrow b\bar{b}ZZ$ \\*
& $\mathtt{SSbbZZ}$ & $X \rightarrow SS \rightarrow b\bar{b}ZZ$ \\*
& $\mathtt{XbbZZ}$  & $X \rightarrow HH/SH/SS \rightarrow b\bar{b}ZZ$ \\

\midrule
\multirow{6}{*}{$b\bar{b}W^+W^-/b\bar{b}ZZ$}
& $\mathtt{HHbbVV}$ & $X \rightarrow HH \rightarrow b\bar{b}W^+W^-/b\bar{b}ZZ$ \\*
& $\mathtt{SbbHVV}$ & $X \rightarrow S(b\bar{b})H(W^+W^-/ZZ)$ \\*
& $\mathtt{SVVHbb}$ & $X \rightarrow S(W^+W^-/ZZ)H(b\bar{b})$ \\*
& $\mathtt{SHbbVV}$ & $X \rightarrow SH \rightarrow b\bar{b}W^+W^-/b\bar{b}ZZ$ \\*
& $\mathtt{SSbbVV}$ & $X \rightarrow SS \rightarrow b\bar{b}W^+W^-/b\bar{b}ZZ$ \\*
& $\mathtt{XbbVV}$  & $X \rightarrow HH/SH/SS \rightarrow b\bar{b}W^+W^-/b\bar{b}ZZ$ \\

\midrule
\multirow{6}{*}{$W^+W^-\tau^+\tau^-$}
& $\mathtt{HHWWtautau}$ & $X \rightarrow HH \rightarrow W^+W^-\tau^+\tau^-$ \\*
& $\mathtt{SWWHtautau}$ & $X \rightarrow S(W^+W^-)H(\tau^+\tau^-)$ \\*
& $\mathtt{StautauHWW}$ & $X \rightarrow S(\tau^+\tau^-)H(W^+W^-)$ \\*
& $\mathtt{SHWWtautau}$ & $X \rightarrow SH \rightarrow W^+W^-\tau^+\tau^-$ \\*
& $\mathtt{SSWWtautau}$ & $X \rightarrow SS \rightarrow W^+W^-\tau^+\tau^-$ \\*
& $\mathtt{XWWtautau}$  & $X \rightarrow HH/SH/SS \rightarrow W^+W^-\tau^+\tau^-$ \\

\midrule
\multirow{6}{*}{$ZZ\tau^+\tau^-$}
& $\mathtt{HHZZtautau}$ & $X \rightarrow HH \rightarrow ZZ\tau^+\tau^-$ \\*
& $\mathtt{SZZHtautau}$ & $X \rightarrow S(ZZ)H(\tau^+\tau^-)$ \\*
& $\mathtt{StautauHZZ}$ & $X \rightarrow S(\tau^+\tau^-)H(ZZ)$ \\*
& $\mathtt{SHZZtautau}$ & $X \rightarrow SH \rightarrow ZZ\tau^+\tau^-$ \\*
& $\mathtt{SSZZtautau}$ & $X \rightarrow SS \rightarrow ZZ\tau^+\tau^-$ \\*
& $\mathtt{XZZtautau}$  & $X \rightarrow HH/SH/SS \rightarrow ZZ\tau^+\tau^-$ \\

\midrule
\multirow{6}{*}{$W^+W^-\tau^+\tau^-/ZZ\tau^+\tau^-$}
& $\mathtt{HHVVtautau}$ & $X \rightarrow HH \rightarrow W^+W^-\tau^+\tau^-/ZZ\tau^+\tau^-$ \\*
& $\mathtt{SVVHtautau}$ & $X \rightarrow S(W^+W^-/ZZ)H(\tau^+\tau^-)$ \\*
& $\mathtt{StautauHVV}$ & $X \rightarrow S(\tau^+\tau^-)H(W^+W^-/ZZ)$ \\*
& $\mathtt{SHVVtautau}$ & $X \rightarrow SH \rightarrow W^+W^-\tau^+\tau^-/ZZ\tau^+\tau^-$ \\*
& $\mathtt{SSVVtautau}$ & $X \rightarrow SS \rightarrow W^+W^-\tau^+\tau^-/ZZ\tau^+\tau^-$ \\*
& $\mathtt{XVVtautau}$  & $X \rightarrow HH/SH/SS \rightarrow W^+W^-\tau^+\tau^-/ZZ\tau^+\tau^-$ \\

\midrule
\multirow{6}{*}{$b\bar{b}\gamma\gamma$}
& $\mathtt{HHbbgamgam}$ & $X \rightarrow HH \rightarrow b\bar{b}\gamma\gamma$ \\*
& $\mathtt{SbbHgamgam}$ & $X \rightarrow S(b\bar{b})H(\gamma\gamma)$ \\*
& $\mathtt{SgamgamHbb}$ & $X \rightarrow S(\gamma\gamma)H(b\bar{b})$ \\*
& $\mathtt{SHbbgamgam}$ & $X \rightarrow SH \rightarrow b\bar{b}\gamma\gamma$ \\*
& $\mathtt{SSbbgamgam}$ & $X \rightarrow SS \rightarrow b\bar{b}\gamma\gamma$ \\*
& $\mathtt{Xbbgamgam}$  & $X \rightarrow HH/SH/SS \rightarrow b\bar{b}\gamma\gamma$ \\

\midrule
\multirow{3}{*}{$t\bar{t}b\bar{b}$}
& $\mathtt{SttHbb}$ & $X \rightarrow S(t\bar{t})H(b\bar{b})$ \\*
& $\mathtt{SSttbb}$ & $X \rightarrow SS \rightarrow t\bar{t}b\bar{b}$ \\*
& $\mathtt{Xttbb}$ & $X \rightarrow S(t\bar{t})H(b\bar{b}) + X \rightarrow SS \rightarrow t\bar{t}b\bar{b}$ \\

\midrule
\multirow{1}{*}{No decay}
& $\mathtt{NoDecay}$ & No scalar decay applied \\

\bottomrule
\end{longtable}

\newpage

\section{Example mass list JSON file}
\label{app:mass_list}

Example~\ref{lst:example-mass-json} shows an example of a \texttt{.json}
file defining a list of mass points from an experimental result.
Each mass list file is contained in the \texttt{data/mass\_points}
directory and uses the naming convention \texttt{<decay>\_<identifier>.json},
i.e., \texttt{SbbHgamgam\_ATLAS.json}, defined using the \texttt{--decay} and
\texttt{--identifier} CLI arguments.

The \texttt{collaboration}, \texttt{arxiv}, \texttt{doi}, \texttt{year}
and \texttt{decay\_channel} fields provide useful analysis metadata.
\texttt{sqrt\_s} provides the center of mass collision energy.
\texttt{units} defines whether the limits for each mass point are
given in \texttt{fb} or \texttt{pb}.
\texttt{includes\_decay} indicates whether the provided limits include
decay branching ratios to SM particles.
The \texttt{mass\_points} block defines the mass points of interest.
\texttt{mX} and \texttt{mS} are the masses of the heavy scalars X and
S, respectively, in GeV.
The limit fields are used to provide the expected and observed experimental
limits.
These are used for the adaptive stopping condition precision and for
the exclusion plot production.
If a limit is unavailable, $-1.0$ can be used as a placeholder with no
impact on the stopping precision or exclusion plotting.

\begin{lstlisting}[language=json,
                   caption={Example mass list configuration using two mass points from~\cite{CMS:2022suh}},
                   label={lst:example-mass-json}]
{
    "collaboration": "CMS",
    "arxiv": "2204.12413",
    "doi": "10.1016/j.physletb.2022.137392",
    "year": 2022,
    "sqrt_s": 13.0,
    "decay_channel": "X->S(bb)H(bb)",
    "units": "fb",
    "includes_decay": true,
    "mass_points": [
        {
            "mX": 900,
            "mS": 60,
            "observed_limit": 18.175,
            "expected_limit": 11.812,
            "expected_limit_m1": 6.794,
            "expected_limit_p1": 22.453,
            "expected_limit_m2": 4.4292,
            "expected_limit_p2": 35.814
        },
        {
            "mX": 900,
            "mS": 70,
            "observed_limit": 5.3782,
            "expected_limit": 5.0312,
            "expected_limit_m1": 3.174,
            "expected_limit_p1": 8.601,
            "expected_limit_m2": 2.2011,
            "expected_limit_p2": 14.828
        }
    ]
}
\end{lstlisting}

\newpage

\bibliographystyle{hunsrt} 
\bibliography{lit.bib}

\end{document}